\documentclass[a4paper]{jpconf}
\usepackage{graphicx}
\usepackage{amsfonts}

\newcommand\beq{\begin{equation}}
\newcommand\eeq{\end{equation}}
\newcommand{\bes}{\begin{eqnarray}}
\newcommand{\ees}{\end{eqnarray}}

\def\vphi{{\varphi}}
\def\extd{\mathrm {d}}
\def\inv{{\mbox{\tiny -1}}}

\newcommand{\SU}{\mathrm{SU}}

\newcommand{\cZ}{{\mathcal Z}}
\newcommand{\cG}{{\mathcal G}}
\newcommand{\cA}{{\mathcal A}}

\newcommand{\cB}{{\mathcal B}}

\newcommand{\cO}{{\mathcal O}}

\newcommand{\one}{\mbox{$1 \hspace{-1.0mm}  {\bf l}$}}

\begin{document}
\title{Singular topologies in the Boulatov model}

\author{Sylvain Carrozza}

\address{Max Planck Institute for Gravitational
Physics, Albert Einstein Institute, Am M\"uhlenberg 1,
14476 Golm, Germany, EU}
\address{Laboratoire de Physique Th\'{e}orique, CNRS UMR 8627,
Universit\'{e} Paris XI, F-91405 Orsay Cedex, France, EU}

\ead{sylvain.carrozza@aei.mpg.de}

\begin{abstract}
Through the question of singular topologies in the Boulatov model, we illustrate and summarize some of the recent advances in Group Field Theory.
\end{abstract}
\section{Introduction}
Group Field Theories\footnote{See \cite{rev} for a recent review.} (GFTs) are genuine quantum field theories defined on group manifolds, especially relevant to quantum gravity in general, and in particular to the question of the continuum limit in Spin Foam models (SF). 
This is mainly because GFTs generate SF models as Feynman amplitudes in their perturbative expansions, thus bringing additional tools and concepts to define and understand the sum over foams.
But interestingly, GFTs can also be seen as a generalization of matrix models \cite{mm} (and a particular class of tensor models), which were particularly successful as discrete models of 2d quantum gravity: a continuum regime could be found that matches exactly the continuum theory. This link with matrix models has recently triggered much progress in GFT and tensor models in general\footnote{See the review \cite{tensor}, and references therein.}. Focusing on the Boulatov model, dual to the Ponzano-Regge model for euclidean 3d quantum gravity, we illustrate some of these results and their consequences for our understanding of singular topologies in GFT.\footnote{The present article is a summary of the motivations and results of the recent work \cite{vertex}, where more technical details can be found.}

\section{Boulatov model and singular topologies}
\subsection{Original model}
The Boulatov model can be defined by the action:
\beq\label{action}
S[\vphi] = \int [\extd g]^{3} \vphi(g_1, g_2, g_3)^{2} + \lambda \int [\extd g]^{6} \vphi(g_1, g_2, g_3) \vphi(g_3, g_5, g_4) \vphi(g_5, g_2, g_6) \vphi(g_4, g_6, g_1) \,,
\eeq
where $\vphi$ is a real field on $\SU(2)^{3}$, required to be invariant under the diagonal action of $\SU(2)$:
\beq\label{gauge}
\forall h \in \SU(2),  \qquad \vphi(hg_1, hg_2, hg_3)  \, = \, \vphi(g_1, g_2, g_3) .
\eeq
This field has the geometrical interpretation of a quantized triangle, the variable $g_i$ being understood as the holonomy encoding parallel transport from the centre of the triangle to the center of the edge $i$. In this interpretation, the interaction term in the action encodes the gluing of four quantum triangles along their boundary edges so as to form a tetrahedron, whereas the (trivial) kinetic term identifies two triangles. When formally expanding the (euclidean) path integral:
\beq
\cZ = \int \extd\mu(\vphi) \, {\rm{e}}^{- S[\vphi]}
\eeq
in powers of $\lambda$, one gets Feynman amplitudes labelled by simplicial complexes. Their building blocks are tetrahedra, to which we associate the kernel of the interaction part of $S$, glued together through their boundary triangles. The propagator (kernel of the kinetic part of the action, supplemented by a projection on gauge invariant fields (\ref{gauge})) encodes these gluings.
Alternatively, these simplicial complexes are represented by dual 3-stranded graphs, where a line of the graph corresponds to a triangle of the simplicial complex, and its three strands are dual to the edges of the same triangle. In a graph $\cG$, one can identify a 2-complex structure (the faces being the closed chains of strands) and show that the GFT amplitude of a simplicial complex is equal to the Ponzano-Regge amplitude of its dual 2-complex:
\beq
\cA^{\cG} = \int [dh_{t}] \prod_{e} \delta\left(H_{e}\right)\,.
\eeq
This formula is written in terms of simplicial data: to a triangle $t$ is associated an holonomy $h_t$ (representing parallel transport along the dual edge in the 2-complex)\footnote{$h_t$ is also the auxiliary variable appearing in the gauge condition (\ref{gauge}).}, $e$ is an edge, and $H_{e}$ is the holonomy around its dual face. It is defined as an ordered product of group elements $h_t$, associated to the triangles which contain the edge $e$:
$
H_{e} = \overrightarrow{\prod_{t \supset e}} h_t \,.
$

\
As in ordinary quantum field theories, the amplitudes are divergent, so the path integral needs to be regulated. This is usually achieved via a regularization of the $\delta$-function on $\SU(2)$, for instance by cutting-off high spins in its Peter-Weyl expansion:
\beq
\delta^{\Lambda}(g) \equiv \sum_{j \in \mathbb{N}/2, j \leq \Lambda} (2 j + 1) \chi^{j}(g)\,,
\eeq 
where $\chi^{j}$ are the characters of $\SU(2)$. Divergences can then be expressed in powers of the value of the regulated $\delta$-function at the identity: $\delta^{\Lambda}(\one) \sim \Lambda^3$.

\
The simplicial complexes generated by the GFT represent (virtual) histories of discrete spaces (here triangulated surfaces), and are the structures from which we hope to recover known (and hopefully unknown) spacetime physics. On the way to the emergence of a continuum spacetime from GFT, with the full structure of differential manifold, it is reasonable to first focus on more primitive aspects, such as topology. Namely, we would like to show that in some regime of the GFT, the simplicial complexes that dominate the perturbative expansion are regular topological manifolds, i.e. topological spaces in which the neighborhood of any point has trivial topology (a point that fails to have this property is called a topological singularity). The problem is that GFTs generically generate very complicated topological spaces\footnote{The topological structure used is the obvious one, induced by the identifications of the boundaries of the tetrahedra, themselves having the topology of a ball in $\mathbb{R}^{3}$.}, with pointlike singularities as well as extended singularities (for example a one dimensional singular subspace) \cite{sing}.

\
Two complementary strategies have been successfully used to control the singular topologies in the Boulatov model, which we describe in the following: the first consists in imposing suitable combinatorial conditions that restrict the class of simplicial complexes summed over in the Feynman expansion; and the second amounts to find scaling bounds showing that the remaining singular complexes are reasonably suppressed with respect to the leading order. 

\subsection{Colored model}
Colored models were originally introduced \cite{color} as a general prescription allowing to restrict, in a controlled way, the class of simplicial complexes generated by GFTs.
In the Boulatov case, it amounts to replacing the real field $\vphi$ by four complex fields $\vphi_\ell$, indexed by a color label $\ell \in \{1, \cdots, 4 \}$. The gauge condition (\ref{gauge}) is unchanged, and the colored action is defined as:
\beq\label{col_action}
S_{{col.}}[\vphi, \overline{\vphi}] = \int |\vphi|^{2} + \lambda \int \vphi_1 \vphi_2 \vphi_3 \vphi_4 + {\rm{c.c.}} \,,
\eeq
where the convolutions of fields follow the same patterns as in the uncolored model (\ref{action}). 
The only effect of the coloring is to reduce the combinatorial complexity of the Feynman graphs generated by the GFT. But, and this is essential, the amplitude of a simplicial complex generated by the colored model is also equal to the Ponzano-Regge partition function. 

\
It can be shown that no simplicial complex with extended singularities is generated by the colored model \cite{color}. There are still singular simplicial complexes appearing in the expansion, but they have only pointlike singularities (located at the vertices): this type of topological spaces, which are regular except on a discrete set, are called pseudomanifolds. Their analysis will be reviewed in the next section.

\
But before that, we need to mention two additional and spectacular properties of the colored model, that are not available in the uncolored version. The first is that the cut-off $\Lambda$ can be used to define a new perturbative expansion \cite{1/N}, exact analog of the $1/N$ expansion in matrix models, that was the essential ingredient missing so far to reproduce their success in higher dimensions. It allowed to prove that, upon a suitable rescaling of the coupling constant $\lambda \rightarrow \lambda / \sqrt{\delta^{\Lambda}(\one)}$, the partition function can be expanded as \cite{critical}:
\beq\label{expansion}
\cZ = [\delta^{\Lambda}(\one)]^{2} \cZ_{0}(\lambda \overline{\lambda}) + \cO([\delta^{\Lambda}(\one)])\, ,
\eeq
where $\cZ_{0}$ contains the contributions of a simple subclass of triangulations of the 3-sphere. Therefore, leading order contributions are not only manifolds, but also have trivial topology. 

\
The second breakthrough \cite{diffeos} (that was again only made possible by the introduction of colors) is the realization of simplicial diffeomorphism symmetry, defined at the level of the amplitudes, as a symmetry of the GFT action itself. Interestingly for us, this symmetry acts on vertices of the triangulations, that is exactly where potential singularities are located. 

\section{Vertex representation}
\subsection{Vertex variables}
As far as the symmetries of the Boulatov model are concerned, the natural variables in a quantum triangle $\vphi(g_1, g_2, g_3)$ are the $G_{ij} \equiv g_i^{\inv} g_j$ \cite{deformed_poincare, diffeos}. It is therefore natural to look for a formulation of the model in which the fields are defined in such variables. Thanks to the gauge invariance (\ref{gauge}), this transformation can actually be performed, and was extensively studied in \cite{vertex}. One obtains a theory with four complex fields $\psi_{\ell}(G_u, G_v, G_w)$, and the same structure of colored action (\ref{col_action}), but with two notable differences: a) the gauge condition (\ref{gauge}) is traded for a closure condition, encoded by a distributional factor $\delta(G_u G_v G_w)$ in the propagator; b) the interaction term encodes the tetrahedral geometry through identifications of vertices common to different triangles (as opposed to edges in the usual formulation), which is reflected in a stranded structure with 3-valent interactions.

\
The main advantage of this formulation is that, for any color $\ell$, it allows to rewrite the amplitudes as integrals of the form:
\beq\label{vertex_amplitude}
\cA^{\cG} \propto  \int [\extd G] \left( \prod_{b \in \cB_{\ell}} \prod_{v \in V_{b}} \delta\left( H_{v, b} \right) \right) \left( \prod_{t_{\ell}} \delta\left( G^{t_\ell}_u G^{t_\ell}_v G^{t_\ell}_w \right)\right)\, ,
\eeq
where: $b \in \cB_{\ell}$ are closed triangulated surfaces around vertices of color $\ell$, called bubbles; $v \in V_{b}$ is a vertex of such a surface; $H_{v, b}$ is the holonomy around the vertex $v$ in $b$; and $t_\ell$ is summed over all the triangles of color $\ell$ in the simplicial complex. 

\
This expression as a very nice geometrical interpretation, describing the effective physics of the field of color $\ell$: the bubble terms represent triangulated boundaries of effective flat cells (first factor), consistently glued through propagators of color $\ell$ (second factor). 
\subsection{Optimal scaling bounds}
In the cut-off theory (and with the scaling of $\lambda$ ensuring the existence of a large $\Lambda$ expansion), the explicit factorization of bubble contributions (\ref{vertex_amplitude}) allows to prove the following scaling bounds:
\beq\label{bound}
\cA^{\cG} = \cO\left([\delta^{\Lambda}(\one)]^{2 - 2 \sum_{b \in \cB_\ell} g_b}\right)\,,
\eeq  
where $g_b$ is the genus of the (closed and orientable) triangulated surface $b$. Remarkably $g_b \in \mathbb{N}$, and $g_b = 0$ if and only if $b$ is a triangulation of the sphere. 
But precisely, a given vertex in the simplicial complex associated to $\cG$ is topologically regular whenever the dual bubble $b$ is spherical. 
So (\ref{bound}) is a bound on singular topologies, that was moreover proven to be optimal \cite{vertex}: for any integers $(g_1, \cdots, g_n)$, there exists a simplicial complex
which has exactly $n$ bubbles of color $\ell$ and genera $(g_1, \cdots, g_n)$, and whose amplitude diverges like $[\delta^{\Lambda}(\one)]^{2 - 2 \sum_{i = 1}^{n} g_i}$. This bound is indeed sharper than what was previously known, since any singular topology will have an amplitude in at most $\cO(1)$, and not only $\cO([\delta^{\Lambda}(\one)])$ as can be deduced from (\ref{expansion}).

\section{Conclusion and outlook}
In this short article, we briefly reviewed the question of singular topologies in the Boulatov model, which is in tension with the discrete spacetime interpretation. This gives an interesting point of view on recent developments in GFT. For instance, a colored model is required both as a way to get rid of very singular simplicial complexes, but also to realize simplicial symmetries as field symmetries, which in turn are essential to understand the left-over singular topologies. The optimal bounds arising from the vertex reformulation also justify by themselves the rescaling of $\lambda$ used in the $\Lambda$ expansion \cite{1/N}: this is the unique scaling for which they can be interpreted as a hierarchical set of bounds that tame the contribution of singularities. Finally, they allow to identify the precise order in $\Lambda$ at which the interpretation of the Boulatov model as a sum over discrete spacetimes ceases to make sense. The good news is that for the leading order (in $[\delta^{\Lambda}(\one)]^{2}$), as well as all corrections in $[\delta^{\Lambda}(\one)]^{\gamma}$ with $\gamma > 0$, this interpretation is available.

\  
This analysis can be generalized to 4d topological models \cite{wip}. Whether this applies also to 4d gravity models needs to be investigated. 


\end{document}